\documentclass[journal]{IEEEtran}

\usepackage{bbding}
\usepackage{threeparttable}
\usepackage{booktabs}
\usepackage{multirow}
\usepackage{makecell}

\usepackage[T1]{fontenc}
\usepackage{graphicx}
\usepackage{bm}
\usepackage[noend]{algpseudocode}
\usepackage{algorithmicx,algorithm}
\usepackage{threeparttable}
\usepackage[sort]{cite}
\usepackage{verbatim}

\usepackage{hyperref}
\hypersetup{hidelinks}

\def\equationautorefname#1#2\null{
	Eq. (#2\null)
}

\ifCLASSOPTIONcompsoc
  \usepackage[caption=false,font=normalsize,labelfont=sf,textfont=sf]{subfig}
\else
  \usepackage[caption=false,font=footnotesize]{subfig}
\fi

\usepackage{amsmath}
\usepackage[cmintegrals]{newtxmath}

\usepackage{stfloats}
\usepackage{xspace}
\usepackage{xcolor}

\begin{document}
%
\title{Alternative Pseudo-Labeling for Semi-Supervised Automatic Speech Recognition}
%
%
%

\author{Han~Zhu,~\IEEEmembership{Student~Member,~IEEE,}
        Dongji~Gao,~\IEEEmembership{Student~Member,~IEEE,}
        Gaofeng~Cheng,~\IEEEmembership{Member,~IEEE,}
        Daniel~Povey,~\IEEEmembership{Fellow,~IEEE,}
        Pengyuan~Zhang,~\IEEEmembership{Member,~IEEE,}
        Yonghong~Yan,~\IEEEmembership{Member,~IEEE,}
\thanks{H. Zhu, G. Cheng, P. Zhang and Y. Yan are with the Key Laboratory of Speech Acoustics and Content Understanding, Institute of Acoustics, Chinese Academy of Sciences, Beijing 100190, China, and also with the University of Chinese Academy of Sciences, Beijing 100049, China (e-mail: zhuhan@hccl.ioa.ac.cn;chenggaofeng@hccl.ioa.ac.cn
;zhangpengyuan@hccl.ioa.ac.cn;yanyonghong@hccl.ioa.ac.cn).}
\thanks{D. Gao is with Johns Hopkins University, Baltimore MD 21218, USA (e-mail: dgao5@jhu.edu).}
\thanks{D. Povey is with the Xiaomi Corporation, Beijing 100085, China (e-mail: dpovey@xiaomi.com).}
}%


%
%

\markboth{Journal of \LaTeX\ Class Files,~Vol.~14, No.~8, August~2015}%
{Shell \MakeLowercase{\textit{et al.}}: Bare Demo of IEEEtran.cls for IEEE Journals}
%



\maketitle
\begin{abstract}

When labeled data is insufficient, semi-supervised learning with the pseudo-labeling technique can significantly improve the performance of automatic speech recognition. However, pseudo-labels are often noisy, containing numerous incorrect tokens. Taking noisy labels as ground-truth in the loss function results in suboptimal performance. Previous works attempted to mitigate this issue by either filtering out the nosiest pseudo-labels or improving the overall quality of pseudo-labels. While these methods are effective to some extent, it is unrealistic to entirely eliminate incorrect tokens in pseudo-labels. In this work, we propose a novel framework named alternative pseudo-labeling to tackle the issue of noisy pseudo-labels from the perspective of the training objective. The framework comprises several components.
Firstly, a generalized CTC loss function is introduced to handle noisy pseudo-labels by accepting alternative tokens in the positions of incorrect tokens. Applying this loss function in pseudo-labeling requires detecting incorrect tokens in the predicted pseudo-labels. In this work, we adopt a confidence-based error detection method that identifies the incorrect tokens by comparing their confidence scores with a given threshold, thus necessitating the confidence score to be discriminative. Hence, the second proposed technique is the contrastive CTC loss function that widens the confidence gap between the correctly and incorrectly predicted tokens, thereby improving the error detection ability. Additionally, obtaining satisfactory performance with confidence-based error detection typically requires extensive threshold tuning. Instead, we propose an automatic thresholding method that uses labeled data as a proxy for determining the threshold, thus saving the pain of manual tuning. Experiments demonstrate that alternative pseudo-labeling outperforms existing pseudo-labeling approaches on datasets in various domains and languages.

\end{abstract}

\begin{IEEEkeywords}
Automatic speech recognition, semi-supervised learning, pseudo-labeling
\end{IEEEkeywords}

%
\IEEEpeerreviewmaketitle

\section{Introduction}

\IEEEPARstart{R}{emarkable} advancements have been accomplished in the field of automatic speech recognition (ASR) \cite{li2021recent,cheng2022eteh}, primarily owing to the advances of deep neural networks~\cite{peddinti2015time,povey2018semi,gulati2020conformer}. Nevertheless, similar to other models based on deep neural networks, the performance of the ASR model largely depends on the quantity of available labeled speech data. Given the high cost associated with human annotation, how to leverage unlabeled data with semi-supervised learning is of great interest in ASR. There are two primary streams of semi-supervised learning approaches: self-supervised learning based model pre-training (SSL)~\cite{baevski2020wav2vec,hsu2021hubert} and pseudo-labeling (PL)~\cite{higuchi2021momentum,park2020improved}.

SSL approaches first pre-train the neural network with unlabeled data, then fine-tune it with labeled data to produce the ASR model.
While SSL can significantly enhance the performance of ASR, several efficiency-related drawbacks still exist. Firstly, SSL bootstraps from random initialization, thus requiring a large number of training updates and computation resources. Secondly, the pre-training and fine-tuning paradigm makes incremental learning~\cite{fu2021incremental} costly. 
Specifically, when we receive a new set of unlabeled data and want to enhance the model, we must resume the pre-training phase to add the new unlabeled data and repeat the fine-tuning phase with all labeled data, thus being expensive in large-scale ASR applications.

PL is widely adopted in ASR due to its simplicity and effectiveness. In typical PL methods, a teacher model trained with available labeled data decodes all unlabeled data. Then both labeled and unlabeled data are used to train the student model with either ground-truth labels or pseudo-labels. PL has some advantages compared with SSL. Firstly, PL starts from an ASR model instead of random initialization, thus being easier to train and requiring less computation resources~\cite{likhomanenko2020slimipl}. Secondly, incremental learning is also convenient in PL since we can directly continue PL on new unlabeled data.

\begin{figure}[t!]
    \centering
	\subfloat[Conventional PL] 
	{ \label{fig:comparison_previous}
		\includegraphics[height=0.6\columnwidth]{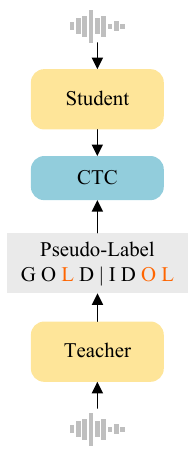}
	}\quad
	\subfloat[Proposed PL] 
	{ \label{fig:comparison_proposed}
		\includegraphics[height=0.6\columnwidth]{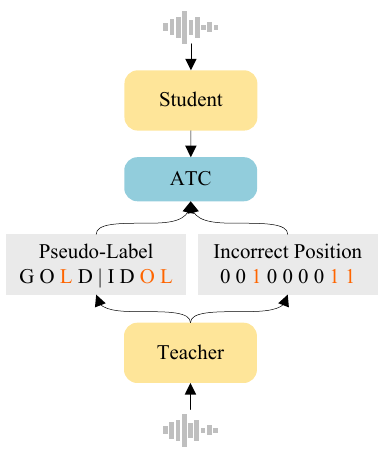}
	}
	\caption{Comparison of conventional and the proposed PL paradigm. The ground-truth label is "good idea", while the predicted pseudo-label is "gold idol". The ATC loss function in the proposed PL paradigm can handle noisy pseudo-labels by leveraging the estimated incorrect positions of pseudo-labels.} 
	\label{fig:comparison}
\end{figure}

However, PL has a troublesome property: pseudo-labels are inevitably erroneous. Fitting the ASR model on incorrect pseudo-label tokens can lead to a significant drop in performance. Several workarounds have been proposed to mitigate the negative effect of noisy pseudo-labels. One intuitive direction is to filter out the noisiest pseudo-labels~\cite{kahn2020self,park2020improved,khurana2021unsupervised}. The typical practice is to use confidence scores to estimate the correctness of pseudo-labels and exclude the low-confident ones. Another line of work focus on directly improving the quality of pseudo-labels. The representative directions include periodically improving the teacher model with the up-to-date student model~\cite{xu2020iterative,manohar2021kaizen,higuchi2021momentum,chen2020semi}, initializing with a better pre-trained model~\cite{zhang2020pushing,zhu2022boosting}, and prediction ensemble~\cite{berthelot2019mixmatch,chen21c_interspeech}.

Although the above approaches can alleviate the negative effect of noisy pseudo-labels to some extent, none can ultimately rule out the errors in pseudo-labels. Therefore, an irrational practice exists in current PL approaches: noisy pseudo-labels are assumed to be ground-truth in the loss function. Consequently, the model is optimized toward the pseudo-labels without doubt, even though we know there are numerous errors in pseudo-labels. In this work, we propose to address the noisy pseudo-label issue from the perspective of the training objective. Specifically, we suggest tackling noisy pseudo-labels by modifying the loss function. Our focus is on the connectionist temporal classification (CTC)~\cite{graves2006connectionist} loss function, which is commonly used in conjunction with PL due to its non-autoregressive property. We introduce a generalized CTC loss function for PL, namely, \textbf{alternative temporal classification (ATC)}, to avoid fitting on the incorrect tokens. In essence, ATC allows accepting alternative tokens on the positions of incorrect tokens in pseudo-labels. 

The essential task to apply ATC in PL is determining which tokens are incorrect in pseudo-labels. Following the mainstream practice, we employ the confidence-based error detection method to determine tokens with lower confidence than a given threshold as incorrect tokens. The quality of confidence estimation determines the ability to detect incorrect tokens, thus influencing the effectiveness of ATC. Hence, we introduce the \textbf{contrastive CTC} loss function to make confidence scores more discriminative between correctly and incorrectly predicted tokens. Moreover, the optimal confidence threshold in confidence-based error detection varies among different datasets, thus requiring manual tuning and imposing additional training burdens. To address this issue, we design an \textbf{automatic thresholding} method that can determine the threshold primarily by referring to the confidence scores of incorrectly predicted tokens in labeled data.

Integrating the aforementioned innovative techniques, including ATC, contrastive CTC, and automatic thresholding, we formulate a novel framework for PL, which we refer to as alternative pseudo-labeling (APL).
We performed detailed experiments on various datasets and showed that APL effectively boosts the performance of ASR and outperforms previous PL-based semi-supervised approaches.

The contributions of this paper are summarized as follows:
\begin{itemize}
\item The overall contribution is a novel PL framework for semi-supervised ASR named APL to address the noisy pseudo-label issue from the training objective perspective.
\item To prevent fitting the model on incorrect tokens, we propose the ATC loss function that can handle noisy pseudo-labels by accepting alternative tokens.
\item To better detect incorrect tokens in pseudo-labels, we present contrastive CTC to widen the confidence gap between correctly and incorrectly predicted tokens.
\item To improve the practicability of APL, we introduce automatic thresholding to avoid manual tuning of thresholds while achieving similar or superior performance.
\end{itemize}

The rest of the paper is organized as follows. In \autoref{sec:releted}, we review related works. Then the proposed approach is introduced in \autoref{sec:proposed}. We describe experimental settings in \autoref{sec:experiment}, and then present experimental results in \autoref{sec:results}. Finally, \autoref{sec:conclusion} concludes the paper.

\section{Related Works}
\label{sec:releted}

In this section, we introduce related works that address the issue of noisy pseudo-labels, which can be summarized into two categories: filtering out noisy pseudo-labels or improving the quality of pseudo-labels. We also review related loss functions designed for noisy labels. 

\subsection{Filter out Noisy Pseudo-labels with Confidence Filtering}

To select better-quality pseudo-labels, some confidence filtering approaches~\cite{zhang2014semi,kahn2020self,park2020improved} utilize decoding-based confidence estimations to select pseudo-labels. However, poorly calibrated neural networks can produce over-confident erroneous predictions and make the decoding-based confidence estimation unreliable \cite{li2021confidence}. To improve the confidence estimation, the model-based approach~\cite{li2021confidence} utilizes an additional confidence estimation module to predict confidence scores. Alternatively, uncertainty ~\cite{khurana2021unsupervised} can also be used as the confidence estimation to filter out the erroneous pseudo-labels, where the uncertainty can be modeled with the prediction variance~\cite{zheng2021rectifying} and estimated via the Monte Carlo dropout~\cite{gal2016dropout}. \cite{zhu2022boosting} further proposes to automatically combine decoding-based and uncertainty-based confidence estimations to achieve better performance. However, relying on confidence filtering approaches to filter all incorrect pseudo-labels is unrealistic since these approaches face a dilemma: keeping fewer pseudo-labels will decrease the available training resource while keeping more pseudo-labels will increase the proportion of incorrect tokens. 

\subsection{Improve the Quality of Pseudo-labels}

Alternatively, some work seeks to improve the overall quality of the pseudo-labels.
IPL~\cite{xu2020iterative} proposes to improve the teacher model by periodically replacing the teacher model with the student model. Continuous PL~\cite{manohar2021kaizen,higuchi2021momentum,chen2020semi} further proposes to continuously improve the teacher model with the up-to-date student model. Self-supervised pre-training~\cite{zhang2020pushing,zhu2022boosting} can also improve the pseudo-label quality by providing a better seed model. Moreover, prediction combination~\cite{berthelot2019mixmatch,chen21c_interspeech} is also shown to be useful in improving the quality of pseudo-labels by fusing multiple predictions. Although helpful to a certain degree, these methods also cannot solve the problem of noisy pseudo-labels since they cannot completely remove incorrect tokens from pseudo-labels.

\subsection{Loss Functions for Noisy Labels}

\begin{figure}[t!]
    \centering
	\includegraphics[width=0.8\columnwidth]{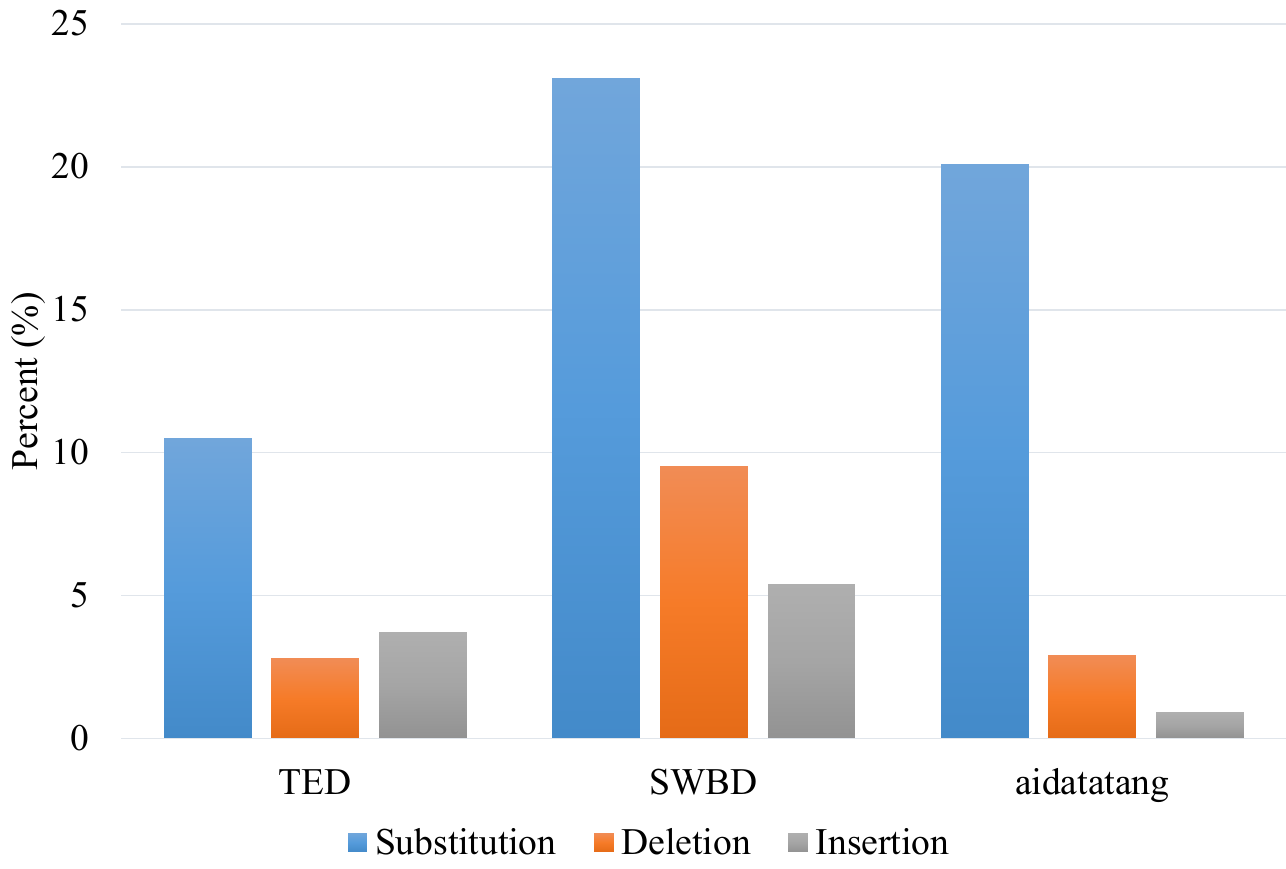}
	\caption{Error types of pseudo-labels in three training sets. The pseudo-labels are generated by a ASR model trained on the LibriSpeech/AISHELL-1 dataset.}
	\label{fig:error_type}
\end{figure}

In order to train the ASR model with noisy labels, a category of loss functions allows training on partial labels. Among them, W-CTC~\cite{cai2021w} allows missing tokens at the beginning or end of the label. Star temporal classification~\cite{pratapstar} further allows an arbitrary number of missing tokens 
anywhere on the label. However, the loss function designed for the partial label is unsuitable in the PL scenario since the pseudo-labels are usually dominated by substitution errors instead of deletion errors (shown in \autoref{fig:error_type}).

To allow arbitrary errors in the label,
Lead2gold~\cite{dufraux2019lead2gold} relies on a noise
model to estimate whether a specific token is correct. However, this noise model is created on a simulation dataset and unavailable in real-world applications. Regarding the PL scenario, N-best lists of pseudo-labels are used instead of 1-best ones in \cite{do2021multiple} and  \cite{moritz2021semi} to reflect the uncertainty of the ASR model, where \cite{do2021multiple} uses the summation of multiple CTC losses as the training objective, and \cite{moritz2021semi} uses a compact lattice-based supervision~\cite{manohar2018semi} to formulate the training objective. Nonetheless, the N-best based loss functions still optimize the model toward the incorrect tokens in pseudo-labels.

\section{Proposed Approach}
\label{sec:proposed}

\subsection{Overview}

\begin{figure*}[!t]
    \centering
	\includegraphics[width=1.5\columnwidth]{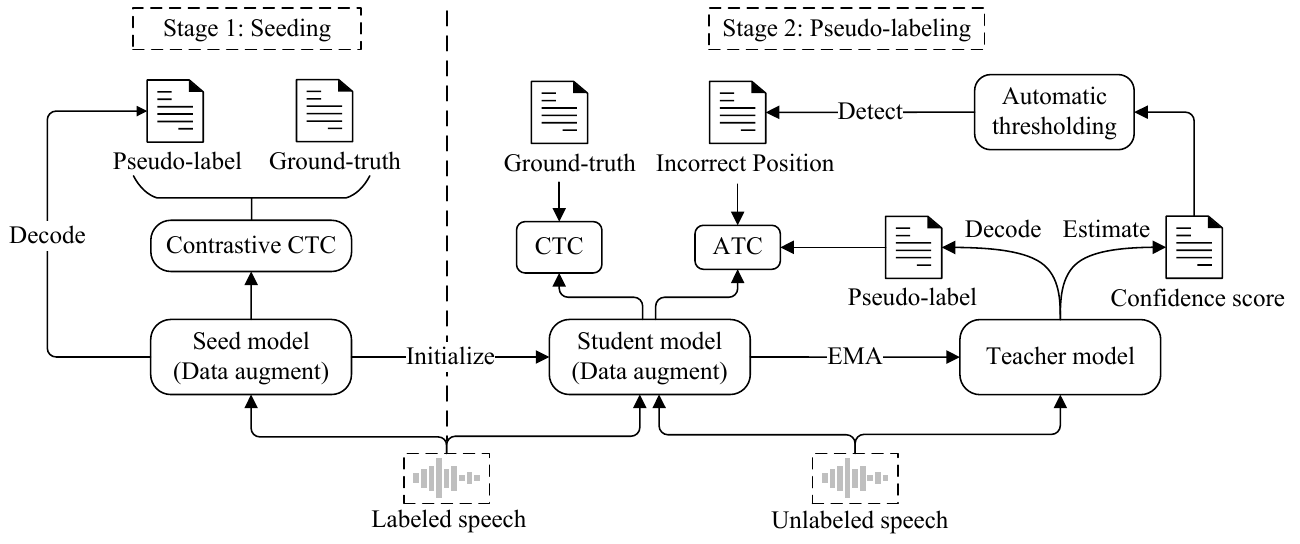}
	\caption{Illustration of APL. Firstly, the seed model is trained on labeled speech with contrastive CTC. Then, initializing from the seed model, the student model is trained on labeled and unlabeled speech with CTC and ATC losses, respectively. Computing the ATC loss requires pseudo-labels and their corresponding positions of incorrect tokens. Pseudo-labels are generated by greedy-decoding of the teacher model, while incorrect tokens are identified as those with confidence scores below the automatic threshold.}
	\label{fig:APL}
\end{figure*}

We illustrate the overall framework of APL in \autoref{fig:APL}. Similar to other PL approaches, APL consists of a seeding stage where the seed ASR model is trained on labeled data and a pseudo-labeling stage where the model is trained on both labeled and pseudo-labeled data. 

In the seeding stage, different from conventional PL approaches, we use \textbf{contrastive CTC} instead of vanilla CTC as the loss function to train the seed ASR model, with the aim of improving the error detection ability. The details of contrastive CTC are introduced in \autoref{sec:contrastive_ctc}.

During the pseudo-labeling stage, we employ the seed model from the previous stage to initialize both the teacher and student models, just like conventional PL approaches. Subsequently, we optimize these models using both labeled and unlabeled data. Leveraging labeled data is straightforward as we can directly train the student model on the labeled data with CTC. What distinguishes our approach from others is the utilization of unlabeled data. Given a batch of unlabeled data, we first use the teacher model to generate their pseudo-labels and corresponding token-level confidence scores. The generation process of token-level confidence score is elaborated in section \autoref{sec:contrastive_ctc}. Then, we pass the confidence scores to the \textbf{automatic thresholding} module (detailed in \autoref{sec:automatic_thresholding}) to detect the incorrect tokens in pseudo-labels. Finally, we update the student model under the \textbf{ATC} loss function with pseudo-labels and positions of incorrect tokens. A detailed description of the ATC can be found in \autoref{sec:ATC}. After each update of the student model, the teacher model is also updated via the exponential moving average (EMA) of the student model, thus continuously improving the quality of pseudo-labels. Concretely,

\begin{equation}
\label{equ:ema_model}
 \xi_{t} = \lambda \xi_{t-1} + (1-\lambda) \theta_{t}  
\end{equation}
where $\xi$ is the teacher model, $\theta$ is the student model and $\lambda \in (0,1)$ is the EMA decay factor.

\subsection{Preliminary: Connectionist Temporal Classification (CTC) and Weighted Finite-State Transducer (WFST)}
\label{sec:CTC}

As ATC is a variant of CTC, we will begin by introducing CTC to clarify our motivations. Additionally, to elucidate the implementation of ATC, we will elaborate on how to realize CTC with WFSTs.

CTC aims to maximize the probability $P(\mathbf{l}|\mathbf{x})$, where $\mathbf{l}=\{l_1, \ldots, l_{U}\}$ is the output sequence (label), $\mathbf{x}=\{\mathbf{x}_1, \ldots, \mathbf{x}_T\}$ is the input sequence, and $U \leq T$. 
Therefore, the CTC loss function to be minimized can be formulated as:

\begin{equation}
\label{equ:ctc}
    L_\text{CTC}(\mathbf{l}, \mathbf{x}) = - \log P(\mathbf{l}|\mathbf{x})
\end{equation}

The computation of $P(\mathbf{l}|\mathbf{x})$ is nontrivial since $\mathbf{l}$ and $\mathbf{x}$ are unaligned. CTC proposes extending the label's length from $U$ to $T$ to construct a path $\mathbf{\pi}$ that satisfies $\mathcal{B}(\mathbf{\pi}) = \mathbf{l}$, where $\mathcal{B}$ is a many-to-one map that removes all blanks $\varnothing$ and the repetitive tokens. For example, $\mathcal{B}($a$\varnothing$ab$\varnothing)=\mathcal{B}(\varnothing$aa$\varnothing \varnothing$abb$)=\ $aab. With this definition, we can compute $P(\mathbf{l} \mid \mathbf{x})$ as the sum of the probabilities of all paths:
\begin{equation}
\label{equ:sum_path}
P(\mathbf{l} \mid \mathbf{x})=\sum_{\mathbf{\pi} \in \mathcal{B}^{-1}(\mathbf{l})} P(\mathbf{\pi} \mid \mathbf{x})
\end{equation}
where $P(\pi \mid \mathbf{x})$ can be computed as:

\begin{equation}
\begin{aligned}
    P(\mathbf{\pi} \mid \mathbf{x}) &= \prod_{t=1}^T P\left(\pi_t \mid \mathbf{\pi}_{1: t-1}, \mathbf{x}\right) \\
    &= \prod_{t=1}^T P\left(\pi_t \mid \mathbf{x}\right) = \prod_{t=1}^T y_{\pi_t}^t 
\end{aligned}
\end{equation}
under the conditional independence assumption of CTC, where $y_{\pi_t}^t$ is the probability of observing label $\pi_t$ at time $t$, which is the output of the CTC model after the softmax operation.

Directly summing over all paths in $\mathcal{B}^{-1}(\mathbf{l})$ is expensive due to the massive number of paths. Alternatively, we can efficiently compute $P(\mathbf{l} \mid \mathbf{x})$ with the forward-backward algorithm. To apply the forward-backward algorithm, we first extend the label sequence $\mathbf{l}$ to $\mathbf{l}^{\prime}$ by inserting blanks between every pair of labels, the start and the end positions. Then, the forward variable $\alpha_{t}(s)$ is defined as the total probability of all partial paths $\mathbf{\pi}_{1: t}$ that end with the label $l_s^{\prime}$: 
\begin{equation}
\alpha_{t}(s)=\sum_{\substack{\pi: \mathcal{B}\left(\mathbf{\pi}_{1: t}\right)=\mathcal{B}\left(\mathbf{1}^{\prime}_{1: s}\right) \\ \pi_t=l_s^{\prime}}} \prod_{t^{\prime}=1}^t y_{\pi_{t^{\prime}}}^{t^{\prime}}
\end{equation}

$\alpha_{t}(s)$ can be recursively computed with the $\alpha$ values from the previous
time $t-1$ following the rules:
\begin{equation}
\label{equ:alpha_rule}
\begin{aligned}
  &\alpha_t(s)=  \\
 &\begin{cases}(\alpha_{t-1}(s)+\alpha_{t-1}(s-1) y_{l_s^{\prime}}^t \enspace \text { if } l_s^{\prime}=\varnothing \text { or } l_{s-2}^{\prime}=l_s^{\prime} \\ \left(\alpha_{t-1}(s)+\alpha_{t-1}(s-1)+\alpha_{t-1}(s-2)\right) y_{l_s^{\prime}}^t \enspace \text { otherwise }\end{cases} 
  \end{aligned}
  \end{equation}

Symmetrically, we can define the backward variable $\beta_t(s)$ and compute it with $\beta$ values from the following time $t+1$:

\begin{equation}
\beta_{t}(s)=\sum_{\substack{\pi: \mathcal{B}\left(\mathbf{\pi}_{t:T}\right)=\mathcal{B}\left(\mathbf{l}^{\prime}_{s: |\mathbf{l}^{\prime}|}\right) \\ \pi_t=l_s^{\prime}}} \prod_{t^{\prime}=t}^T y_{\pi_{t^{\prime}}}^{t^{\prime}}
\end{equation}

\begin{equation}
\label{equ:beta_rule}
\begin{aligned}
  &\beta_t(s)=  \\
 &\begin{cases}(\beta_{t+1}(s)+\beta_{t+1}(s+1) y_{l_s^{\prime}}^t \enspace \text { if } l_s^{\prime}=\varnothing \text { or } l_{s+2}^{\prime}=l_s^{\prime} \\ \left(\beta_{t+1}(s)+\beta_{t+1}(s+1)+\beta_{t+1}(s+2)\right) y_{l_s^{\prime}}^t \enspace \text { otherwise }\end{cases} 
  \end{aligned}
  \end{equation}

Finally, we can compute $P(\mathbf{l} \mid \mathbf{x})$ with forward and backward variables as:

\begin{equation}
P(\mathbf{l} \mid \mathbf{x})= \sum_{t=1}^{T} \sum_{s=1}^{|\mathbf{1^{\prime}}|} \frac{\alpha_t(s) \beta_t(s)}{y_{l^{\prime}_s}^t}
\end{equation}

The above computation procedure of CTC loss can be efficiently implemented with the differentiable WFST framework~\cite{mohri2002weighted,hannun2020differentiable}. A WFST is a finite automaton where each arc maps an input label to an output label with a specific weight. When no output label exists, WFST degenerates to WFSA (weighted finite-state acceptor). In this work, we also refer to WFSA as WFST for simplicity.
To compute the CTC loss with differentiable WFST, firstly, we represent the recursive rules in \autoref{equ:alpha_rule} and \autoref{equ:beta_rule} with a WFST label graph shown in \autoref{fig:CTC}~\cite{miao2015eesen,laptev22_interspeech}. Then, we compose this label graph with an emission graph that represents the output of the neural network. Finally, the CTC loss is computed as the total score of the composed graph with log semiring. With this implementation, we can flexibly modify the CTC loss function for our purpose by manipulating the WFST label graph.

\subsection{Alternative Temporal Classification (ATC) }
\label{sec:ATC}

CTC works well when the ground-truth label $\mathbf{l}$ is provided. However, when only a partially correct label is available, e.g., pseudo-label $\mathbf{l}^{\prime}$, the model optimized with the CTC loss $L_\text{CTC}(\mathbf{l}^{\prime}, \mathbf{x})$ will mistakenly increase the probabilities of incorrect tokens. In order to alleviate this issue, we propose to accept alternative tokens in the positions of incorrect tokens, which formulate an adaptation of CTC, i.e. alternative temporal classification (ATC).

\begin{figure*}[!t]
    \centering
	\subfloat[CTC] 
	{ \label{fig:CTC}
		\includegraphics[width=1.3\columnwidth]{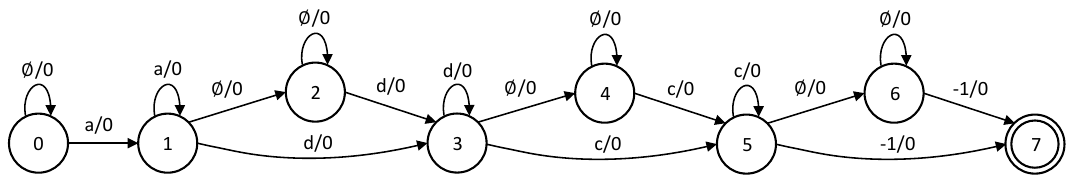}
	}
	
	\subfloat[ATC-R] 
	{ \label{fig:ATC_replace_arc}
		\includegraphics[width=1.3\columnwidth]{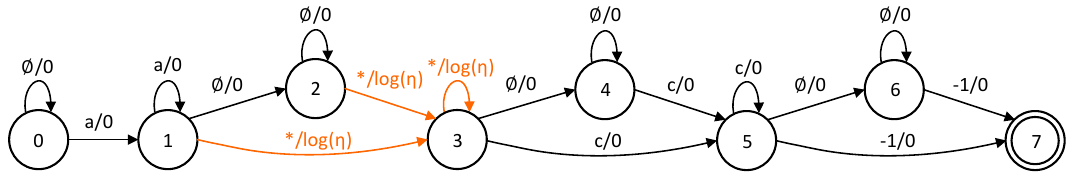}
	}
 
	\subfloat[ATC-A] 
	{ \label{fig:ATC_add_arc}
		\includegraphics[width=1.3\columnwidth]{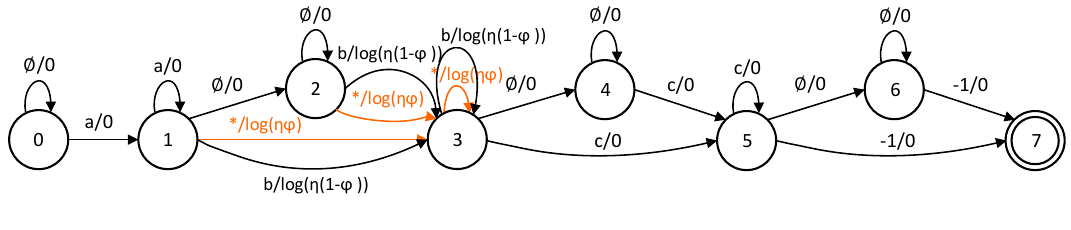}
	}
	\caption{Label graphs of CTC and ATC. ATC-R and ATC-A are variants of ATC that replace or add arcs to the CTC label graph. The ground-truth is "abc", while the pseudo-label is “adc”, i.e., "d" is an incorrect token. $\varnothing$ is the blank token. $*$ denotes this arc accepts all tokens but $\varnothing$. Arcs with "-1" are intentionally designed in k2 to point to the final state.}
	\label{fig:CTC_ATC}
\end{figure*}

Suppose the pseudo-label token $\mathbf{\pi}^{\prime}_{t}$ in the path $\mathbf{\pi}^{\prime} \in \mathcal{B}^{-1}(\mathbf{l}^{\prime})$ is an error, the general form of ATC is to replace the probability $y^{t}_{\pi^{\prime}_{t}}$ as a weighted sum of probabilities of all tokens, except for the blank token, in the loss function.
\begin{equation}
    \hat{y}^{t}_{\pi^{\prime}_{t}} = \sum_{i \neq \varnothing} y^{t}_{i} b(i \mid \pi^{\prime}_{t})
\end{equation}
where the weight $b(i \mid \pi^{\prime}_{t})$ is the benefit of using $i$ as the alternative token for $\pi^{\prime}_{t}$. 

Concretely, we propose two variants of ATC, namely ATC with replaced arc (ATC-R) and ATC with added arc (ATC-A). 
We first introduce ATC-R as follows. When no extra information is available, we can assume that every token except blank has an equal chance of being the alternative token of $\pi^{\prime}_{t}$, i.e., $b(i \mid \pi^{\prime}_{t})$ is a constant for all tokens $i \neq \varnothing$. Therefore, the alternative probability becomes:
\begin{equation}
\label{equ:replace_arc}
    \hat{y}^{t}_{\pi^{\prime}_{t}} = \eta \sum_{i \neq \varnothing} y^{t}_{i} = \eta y^{t}_{*}
\end{equation}
where $*$ is a new token whose probability is the summation of all tokens except the blank. 

Note that the above alternative probability introduces a scale factor $\eta \in (0,1)$ to scale down the value. This is due to the fact that the probability of $*$ is larger compared with other tokens. As a result, after the replacement, the paths with more frames aligned to the $*$ token will have a larger gradient, encouraging the model to produce the $*$ token instead of real tokens. The introduced scale factor can effectively circumvent this effect by scaling down the corresponding gradients.

Formally, we perform analysis by comparing the desired gradient and ATC-R's gradient. 
Suppose $\hat{\mathbf{l}}$ is the label after replacing all incorrect tokens in pseudo-label $\mathbf{l}^{\prime}$ with correct ones. For probability $P(\mathbf{\pi} \mid \mathbf{x})$ of any path $\mathbf{\pi} \in \mathcal{B}^{-1}(\hat{\mathbf{l}})$, its gradient with respect to $y^t_{\pi_t}$ is:

\begin{equation}
\label{equ:gradient_desired}
\frac{\partial P(\mathbf{\pi} \mid \mathbf{x})}{\partial y^t_{\pi_t}}=\frac{\partial \prod_{t^{\prime}=1}^T y_{\pi_{t^{\prime}}}^{t^{\prime}}}{\partial y^t_{\pi_t}}=
\left(\prod_{t^{\prime}=1}^{t-1} y_{\pi_{t^{\prime}}}^{t^{\prime}}\right) \cdot\left(\prod_{t^{\prime}=t+1}^T y_{\pi_{t^{\prime}}}^{t^{\prime}}\right) 
\end{equation}

Similarly, in ATC-R, the gradient of $P(\mathbf{\pi} \mid \mathbf{x})$ with respect to the probability $\hat{y}^{t}_{\pi_{t}}$ would be:

\begin{equation}
\label{equ:gradient_atc}
\frac{\partial P(\mathbf{\pi} \mid \mathbf{x})}{\partial \hat{y}^t_{\pi_t}}=\frac{\partial \prod_{t^{\prime}=1}^T \hat{y}_{\pi_{t^{\prime}}}^{t^{\prime}}}{\partial \hat{y}^t_{\pi_t}}=
\left(\prod_{t^{\prime}=1}^{t-1} \hat{y}_{\pi_{t^{\prime}}}^{t^{\prime}}\right) \cdot\left(\prod_{t^{\prime}=t+1}^T \hat{y}_{\pi_{t^{\prime}}}^{t^{\prime}}\right) 
\end{equation}
where 
\begin{equation}
\hat{y}^{t}_{\pi_{t}} = 
\begin{cases} 
\eta y^{t}_{*} \enspace \text { if } \pi_{t} \neq \pi^{\prime}_{t}   \\
y^{t}_{\pi_{t}} \enspace \text { otherwise }
\end{cases} 
\end{equation}

Dividing \autoref{equ:gradient_atc} with \autoref{equ:gradient_desired}, we get the ratio of the ATC-R's gradient to the desired gradient:

\begin{equation}
\label{equ:divide}
    \frac{\partial P(\mathbf{\pi} \mid \mathbf{x})}{\partial \hat{y}^t_{\pi_t}} / \frac{\partial P(\mathbf{\pi} \mid \mathbf{x})}{\partial y^t_{\pi_t}} = \prod_{t: \pi_{t} \neq \pi^{\prime}_{t}} \eta \frac{y^{t}_{*}}{y^{t}_{\pi_{t}}} 
\end{equation}
where $\frac{y^{t}_{*}}{y^{t}_{\pi_{t}}} $ is always larger than 1.

Consequently, without the scale factor $\eta$, the above ratio will be proportional to the number of frames aligned to $*$ in the current path. ATC-R will have a property that the paths with more frames aligned to the $*$ token will receive a larger gradient, thus being more encouraged. The scale factor $\eta$ can be used to neutralize this effect.

Like CTC, the computation of the ATC-R loss is also implemented with differentiable WFST. The differences between WFST label graphs of CTC and ATC-R are described as follows. Firstly, to substitute all incorrect tokens’ probabilities with the probability of $*$, we replace the arc of the incorrect token in \autoref{fig:CTC} with an arc of $*$, as illustrated in \autoref{fig:ATC_replace_arc}.
Then, the scale factor $\eta$ can be incorporated by adding a penalty of $\log (\eta)$ on the arcs of $*$. 

There is one potential concern with the ATC-R approach in real-world scenarios: the detection of incorrect tokens is imperfect, resulting in a number of false alarm detection, i.e., detecting correct tokens as incorrect. Replacing all detected tokens with $*$ will ignore this fact and treat all tokens equally. To mitigate this issue, we propose another variant of ATC: ATC-A. Unlike ATC-R, ATC-A revises the alternative probability to favors the original token, thereby enhancing the system's robustness against the false alarm detection of incorrect tokens. Specifically, we can modify \autoref{equ:replace_arc} as:
\begin{equation}
\label{equ:add_arc}
    \hat{y}^{t}_{\pi_{t}} = \eta (\psi y^{t}_{*} + (1-\psi) y^{t}_{\pi_{t}})
\end{equation}
where $\psi \in (0, 1)$ is the trade-off hyper-parameter.

Similarly, as shown in \autoref{fig:ATC_add_arc}, ATC-A can be implemented by adding an arc of $*$ in parallel with the arcs of incorrect tokens. Penalties $\log (\eta \psi)$ and $\log (\eta (1-\psi))$ are added on the arc of $*$ and incorrect tokens, respectively.
\subsection{Contrastive CTC for better confidence estimation}
\label{sec:contrastive_ctc}

To detect incorrect tokens in predicted pseudo-labels, we use the CTC-based token-level confidence score as the measurement of the correctness of tokens. And the tokens with confidence less than a given threshold will be detected as incorrect tokens.

The computation of this confidence score is related to the greedy-decoding operations of CTC models. Given the probability of each token from the neural network, greedy-decoding has two operations: merging consecutive identical tokens and removing blank tokens. 
Similarly, CTC-based token-level confidence scores are computed by averaging the probabilities of these consecutive identical tokens. An alternative way is to use the maximum probability of these consecutive identical tokens as the confidence score. In practice, the average and the maximum implementations lead to similar results. 

The quality of confidence estimation determines the ability to detect incorrect tokens and affects the effectiveness of ATC. Therefore, we propose a contrastive training method to improve confidence estimation.

Training with CTC loss would optimize the predictions toward the ground-truth paths, thus increasing the confidence of the ground-truth tokens. However, the desired confidence score should not only be higher on the ground-truth tokens but also be lower for other tokens. Lacking such constraints in the training objective could lead to over-confident predictions of incorrect tokens~\cite{li2021confidence} and make confidence-based error detection difficult. Our proposal to address this issue is to use contrastive learning on the labeled data to make the confidence score more discriminative. 

Suppose the ground-truth label is $\mathbf{l}$, the input sequence is $\mathbf{x}$, the contrastive CTC loss function can be formulated as:
\begin{equation}
\label{equ:contrastive_ctc}
L_\text{c}=L_\text{CTC}(\mathbf{l}, \mathbf{x}^{\prime})-\gamma L_\text{CTC}(\mathbf{l}^{\prime}, \mathbf{x}^{\prime})
\end{equation}
where $\mathbf{x}^{\prime}$ is the augmented version of $\mathbf{x}$. $\mathbf{l}^{\prime}$ is the greedy-decoding result of $\mathbf{x}^{\prime}$ with dropout enabled so that $\mathbf{l}^{\prime}$ contains many errors. And $\gamma \in (0,1)$ is a scale factor. 

The proposed contrastive CTC loss function will not only increase the probabilities of ground-truth tokens but also decrease the probabilities of incorrectly predicted tokens. We give the analysis as follows.

Combining \autoref{equ:contrastive_ctc} with \autoref{equ:ctc}, we get:

\begin{equation}
    L_\text{c} = - \log P(\mathbf{l}|\mathbf{x}^{\prime}) + \gamma \log P(\mathbf{l}^{\prime}|\hat{\mathbf{x}})
\end{equation}

According to \autoref{equ:sum_path}, the above equation becomes:

\begin{equation}
    \label{equ:contrastive_ctc_extend}
    L_\text{c} = - \log \sum_{\mathbf{\pi} \in \mathcal{B}^{-1}(\mathbf{l})} P(\mathbf{\pi} \mid \mathbf{x}^{\prime}) + \gamma \log \sum_{\mathbf{\pi}^{\prime} \in \mathcal{B}^{-1}(\mathbf{l}^{\prime})} P(\mathbf{\pi}^{\prime} \mid \mathbf{x}^{\prime})
\end{equation}

For simplicity, we consider there are only substitution errors in $\mathbf{l}^{\prime}$,  and there are no consecutive identical tokens in $\mathbf{l}$ and $\mathbf{l}^{\prime}$. In this case, the set of paths $\mathcal{B}^{-1}(\mathbf{l})$ should have a one-to-one map with $\mathcal{B}^{-1}(\mathbf{l}^{\prime})$. 

In the extreme case where all corresponding paths $\mathbf{\pi}$ and $\mathbf{\pi}^{\prime}$ in $\mathcal{B}^{-1}(\mathbf{l})$ and $\mathcal{B}^{-1}(\mathbf{l}^{\prime})$ are the same, i.e., decoding results are the same with ground-truth labels, the contrastive loss degenerates to $(1-\gamma) L_\text{CTC}(\mathbf{l}, \mathbf{x}^{\prime})$ and has the same optimization direction of vanilla CTC, i.e., increasing the probabilities of ground-truth tokens.

In normal cases where errors exist in $\mathbf{l}^{\prime}$, the corresponding paths $\mathbf{\pi}^{\prime}$ will also contain such errors. Therefore, the second term in \autoref{equ:contrastive_ctc_extend} will also decrease the probabilities of incorrect tokens in path $\mathbf{\pi}$. 

\begin{figure}[t!]
    \centering
	\includegraphics[width=0.7\columnwidth]{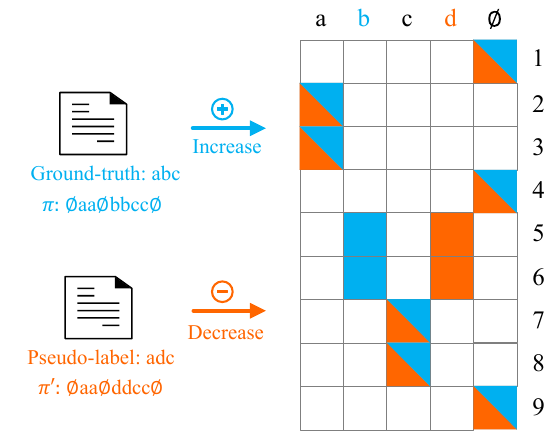}
	\caption{Illustration of contrastive CTC loss function. Probabilities in the path of the ground-truth label (blue color) are increased, while the ones in the path of the pseudo-label (orange color) are decreased.}
	\label{fig:contrastive_CTC}
\end{figure}

For example, as shown in \autoref{fig:contrastive_CTC}, suppose the ground-truth label $\mathbf{l}$ is "abc", and the pseudo-label $\mathbf{l}^{\prime}$ is "adc", i.e., "b" is incorrectly predicted as "d". We take a path $\mathbf{\pi}=\ $"$\varnothing$aa$\varnothing$bbcc$\varnothing$" from $\mathcal{B}^{-1}(\mathbf{l}^{\prime})$. Then, the corresponding path in $\mathcal{B}^{-1}(\mathbf{l})$ is $\mathbf{\pi}^{\prime}=\ $"$\varnothing$aa$\varnothing$ddcc$\varnothing$". According to \autoref{equ:contrastive_ctc_extend}, the probabilities $y^{t}_{\pi_{t}}$ will be increased while the probabilities $y^{t}_{\pi^{\prime}_{t}}$ will be decreased. Consequently, probabilities $\{y^{5}_{b}, y^{6}_{b}\}$ will be increased and probabilities $\{y^{5}_{d}, y^{6}_{d}\}$ will be decreased. The shared probabilities, i.e., $\{y^{1}_{a}, y^{2}_{a}, \ldots\}$, will also be increased since there is a scale factor $\gamma \in (0,1)$.

In conclusion, contrastive CTC would result in an amplified confidence gap between correctly and incorrectly predicted tokens. Consequently, it would be easier to identify incorrect tokens with confidence scores.

\subsection{Automatic thresholding}
\label{sec:automatic_thresholding}

Confidence-based error detection methods require a confidence threshold to determine incorrect tokens. However, we can not simply fix the threshold since it can be model-dependent and data-dependent. And manually tuning the threshold is expensive since we need to train and evaluate the model multiple times to determine the optimal threshold. To address this issue, we design an automatic thresholding method that can dynamically determine the threshold according to training status.

The main idea is to use the labeled data as the proxy to determine the threshold of the unlabeled data. Since the ground-truth labels of labeled data are available, we can measure the average confidence of incorrectly predicted tokens in labeled data. This average confidence level is indicative of the model's confidence when making incorrect predictions, and as such, can function as a suitable threshold for unlabeled data. 

The average confidence is subject to change over time as models tend to exhibit higher confidence scores with longer training updates. In order to account for this dynamic fluctuation in confidence, we employ the exponential moving average (EMA) technique to compute the dynamic average confidence~\cite{wang2022freematch}:

\begin{equation}
T_t^e=(1-\lambda) C_t^e+\lambda T_{t-1}^e
\end{equation}
where $C_t^e$ is the average confidence of incorrectly predicted tokens of the labeled data in each update, and $\lambda \in (0,1)$ is the EMA decay factor. Note that we use the same $\lambda$ for the EMA update of the model parameters in \autoref{equ:ema_model} and here.

The average confidence can be distinct for labeled and unlabeled data, especially when they are in different domains. In order to bridge the gap of confidence between the unlabeled and labeled data, we propose the relative correction method. Specifically, we compute the dynamic average confidence of unlabeled and labeled data with the EMA technique, which are denoted as $T_t^u$ and $T_t^l$, respectively:
\begin{equation}
T_t^u=(1-\lambda) C_t^u+\lambda T_{t-1}^u
\end{equation}
\begin{equation}
T_t^l=(1-\lambda) C_t^l+\lambda T_{t-1}^l
\end{equation}
where $C_t^u$ and $C_t^l$ are average confidences of all unlabeled and labeled data, respectively.
Finally, the confidence threshold with the relative correction is computed as:
\begin{equation}
T_t=\frac{T_t^u}{T_t^l} T_t^e
\end{equation}

This confidence threshold is dynamically computed during training according to training status. Therefore, it can adapt to different data conditions.

\begin{figure}[t!]
    \centering
	\includegraphics[width=1.0\columnwidth]{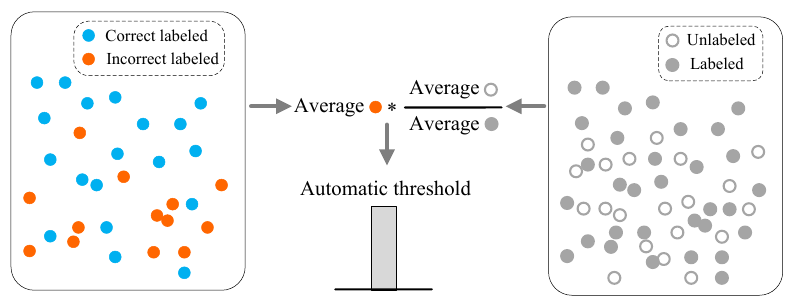}
	\caption{Illustration of automatic threshold. The average confidence of incorrect predictions of the labeled data is used as the base threshold, which is then corrected by the average confidence ratio between unlabeled and labeled data.}
	\label{fig:automatic_threshold}
\end{figure}

\section{Experimental Setup}
\label{sec:experiment}
\subsection{Corpus}

In order to test the effectiveness and generalization ability of the proposed approach, we prepare experimental corpora in two languages: English and Chinese. For English experiments, LibriSpeech~\cite{panayotov2015librispeech} train-clean-100h set is as the labeled dataset, TED-LIUM v3 (TED) \cite{hernandez2018ted} and SwitchBoard (SWBD) \cite{godfrey1993switchboard} are used as unlabeled datasets. Note that TED is slightly mismatched with labeled data (Librispeech), and SWBD is severely mismatched with labeled data. For Chinese experiments, we use AISHELL-1~\cite{aishell_2017} as the labeled data and aidatatang\_200zh (aidatatang)~\cite{aidatatang} as the unlabeled data.
For TED and aidatatang, standard train/dev/test splits are used. In terms of SWBD, the development set is RT-03S \cite{rt03}, and the testing sets are Hub05 Eval2000 \cite{eval2000} SwitchBoard (H-SB) and CallHome (H-CH) subsets. All audios are re-sampled to 16kHz, and transcripts are pre-processed to upper-case letters with no punctuation except apostrophes. The structure of all datasets is shown in \autoref{tab:datasets}. 

\begin{table}[htbp]
  \centering
  \caption{The structure of labeled and unlabeled datasets (hours).}
    \begin{tabular}{lccccc}
    \toprule
    \multirow{3}[6]{*}{Split} & \multicolumn{2}{c}{Labeled data} & \multicolumn{3}{c}{Unlabeled data} \\
\cmidrule(lr){2-3} \cmidrule(lr){4-6}          & English & Chinese & \multicolumn{2}{c}{English} & Chinese \\
\cmidrule(lr){2-2} \cmidrule(lr){3-3} \cmidrule(lr){4-5} \cmidrule(lr){6-6}  & LibriSpeech & AISHELL-1 & TED   & SWBD  & aidatatang \\
    \midrule
    Train & 101   & 151   & 452   & 319   & 140 \\
    Dev   & 5.1   & 18    & 2     & 6     & 20 \\
    Test  & 5.3   & 10    & 3     & 4     & 40 \\
    \bottomrule
    \end{tabular}%
  \label{tab:datasets}%
\end{table}%

\subsection{Implementation Details}
The ATC loss function is implemented with the k2 toolkit\footnote{\url{https://github.com/k2-fsa/k2}}, and the training procedure is implemented with the FAIRSEQ \cite{ott2019fairseq} toolkit. 
We adopt the model architecture of wav2vec 2.0~\cite{baevski2020wav2vec} as the backbone in our experiment, and all models are initialized with the base wav2vec 2.0 model trained on LibriSpeech to speed up convergence.

During the seeding stage, the effective batch size is 25.6m samples, and the seed model is trained with labeled data using contrastive CTC for 20k updates. Then, during the pseudo-labeling stage, the model is trained with both labeled and unlabeled data for another 20k updates, where the first 10k updates are trained with ATC, and the last 10k updates are trained with CTC. The effective batch size for labeled and unlabeled data in each update are both 12.8m samples. The data augmentation strategy for both stages follows \cite{baevski2020wav2vec}, i.e., masking in both time and channel dimensions, similar to SpecAugment~\cite{park2019specaugment}. Although there are several hyper-parameters in APL, we use a fixed set of hyper-parameters to show that it generalizes well on different datasets. The hyper-parameters are: $\lambda=0.999$, $\eta=0.3$, $\gamma=0.5$.
As for evaluation, we use beam-search decoding with the 4-gram LM, where the LM is trained with the corresponding labeled dataset. The teacher model is used for evaluation since it provides better performance.

\section{Results}
\label{sec:results}

\subsection{Comparison with Previous Methods}
\begin{table*}[htbp]
  \centering
  \caption{Results on various datasets.}
    \begin{tabular}{l|cc|cc|ccc|ccc|cc|cccc}
    \toprule
    \multicolumn{1}{l}{\multirow{3}[6]{*}{Method}} & \multicolumn{4}{c}{TED}       & \multicolumn{6}{c}{SWBD}                      & \multicolumn{4}{c}{aidatatang} & \multicolumn{2}{c}{AVG} \\
\cmidrule(lr){2-5}  \cmidrule(lr){6-11}  \cmidrule(lr){12-15}  \cmidrule(lr){16-17}  \multicolumn{1}{l}{} & \multicolumn{2}{c}{w/o LM} & \multicolumn{2}{c}{w/ LM} & \multicolumn{3}{c}{w/o LM} & \multicolumn{3}{c}{w/ LM} & \multicolumn{2}{c}{w/o LM} & \multicolumn{2}{c}{w/ LM} & \multirow{2}[4]{*}{w/o LM} & \multirow{2}[4]{*}{w/ LM} \\
\cmidrule(lr){2-3} \cmidrule(lr){4-5}  \cmidrule(lr){6-8} \cmidrule(lr){9-11} \cmidrule(lr){12-13} \cmidrule(lr){14-15}    \multicolumn{1}{l}{} & dev   & \multicolumn{1}{c}{test} & dev   & \multicolumn{1}{c}{test} & RT03  & H-SB  & \multicolumn{1}{c}{H-CH} & RT03  & H-SB  & \multicolumn{1}{c}{H-CH} & dev   & \multicolumn{1}{c}{test} & dev   & test  &       &  \\
    \midrule
    \multicolumn{17}{l}{\textit{Supervised-only}} \\
    CTC & 19.3  & 19.8  & 14.0  & 14.0  & 46.5  & 35.3  & 43.2  & 38.2  & 27.7  & 35.8  & 26.9  & 27.4  & 23.4  & 24.0  & 31.2  & 25.3 \\
    \midrule
    \multicolumn{17}{l}{\textit{Semi-supervised}} \\
    PL    & 13.0  & 12.7  & 11.1  & 10.9  & 37.2  & 26.4  & 34.1  & 32.4  & 22.2  & 29.9  & 18.7  & 19.3  & 18.3  & 19.0  & 23.1  & 20.5 \\
    IPL~\cite{xu2020iterative}   & \textbf{11.5} & 11.7  & \textbf{10.6} & 10.8  & 30.4  & 20.2  & 28.5  & 28.5  & 18.6  & 26.8  & 15.0 & 15.8 & 16.4 & 17.2 & 19.0 & 18.4 \\
    MPL~\cite{higuchi2021momentum}   & 12.8  & 11.3  & 11.4  & 9.7   & 24.8  & 18.3  & 24.4  & 20.7  & 14.7  & 20.5  & 16.1  & 16.5  & 14.2  & 15.0  & 17.7  & 15.2 \\
    APL   & 12.2  & \textbf{11.3} & 10.7  & \textbf{9.2} & \textbf{22.0} & \textbf{16.0} & \textbf{22.2} & \textbf{18.5} & \textbf{13.3} & \textbf{19.1} & \textbf{14.8} & \textbf{15.4} & \textbf{13.5} & \textbf{14.3} & \textbf{16.3} & \textbf{14.1} \\
    \bottomrule
    \end{tabular}%
  \label{tab:main}%
\end{table*}%

We compare APL with four baseline methods:

\begin{itemize}
\item \emph{CTC}: Supervised-only baseline trained on all labeled data with CTC for 20k updates. The following semi-supervised baselines all use this model as the seed.
\item \emph{PL}: Vanilla pseudo-labeling method that first pseudo-labels all unlabeled data with the supervised-only seed model and then trains the seed model on both labeled and unlabeled data for 20k updates. A 4-gram LM in the domain of the corresponding labeled data is used to generate pseudo-labels. The averaging of the last three checkpoints is used for evaluation.
\item \emph{IPL}: Iterative pseudo-labeling (IPL)\cite{xu2020iterative} applies the PL procedure for multiple iterations to iterative refine the pseudo-labels. Specifically, we perform the PL process for four iterations, each consisting of 5k updates. The last three checkpoints in each iteration are averaged for pseudo-labeling and evaluation.
\item \emph{MPL}: Momentum pseudo-labeling (MPL)~\cite{higuchi2021momentum} is a continuous pseudo-labeling method that utilizes the EMA version of student model as the teacher model to generate pseudo-labels so that pseudo-labels are continually improved. The total number of training updates is 20k, and the EMA decay factor is 0.999. As for evaluation, MPL also uses the teacher model for better performance.
\end{itemize}

\autoref{tab:main} demonstrate the performances of APL and four baselines on various conditions: English dataset with a minor mismatch between unlabeled and labeled data (TED), English dataset with a significant mismatch between unlabeled and labeled data (SWBD) and Chinese dataset (aidatatang). 

Experimental results show that all semi-supervised methods can significantly outperform the supervised-only baseline. Among semi-supervised methods, IPL consistently outperforms PL by iteratively improving the teacher model. Comparing IPL and MPL, IPL lags behind MPL in most datasets since the IPL's refinement frequency of the teacher model is much lower than MPL. Note that IPL outperforms MPL on the development set of the TED dataset. Since IPL utilizes an LM in the domain of labeled data while MPL does not, when evaluated on the TED dataset whose linguistic style is similar to the labeled data (LibriSpeech), IPL has an advantage by improving the quality of pseudo-labels with the LM.

Comparing APL with other approaches, APL consistently outperforms MPL by replacing the loss function from CTC to ATC. APL also outperforms IPL on most datasets. The only exception is the w/o LM result on the development of the TED dataset, for the reason we explained above. Nonetheless, when evaluated w/ LM, the overall performance of APL on the TED dataset is still clearly better than IPL. 

Utterance-level filtering is a widely adopted method to address the noisy pseudo-label issue. Therefore, we also compare the proposed APL approach with the MPL approach equipped with utterance-level filtering in \autoref{tab:utt_filter}. 

\begin{table}[htbp]
  \centering
  \caption{Comparison of utterance-level filtering and APL}
    \begin{tabular}{l|cc|cc}
    \toprule
    \multicolumn{1}{l}{\multirow{3}[6]{*}{Method}} & \multicolumn{4}{c}{TED} \\
\cmidrule{2-5}    \multicolumn{1}{l}{} & \multicolumn{2}{c}{w/o LM} & \multicolumn{2}{c}{w/ LM} \\
\cmidrule(lr){2-3} \cmidrule(lr){4-5}    \multicolumn{1}{l}{} & dev   & \multicolumn{1}{c}{test} & dev   & test \\
    \midrule
    MPL   & 12.8  & 11.3  & 11.4  & 9.7 \\
    MPL+utt. filter (T=0.8) & 12.9  & 11.4  & 11.3  & 9.6 \\
    MPL+utt. filter (T=0.9) & 12.7  & 11.9  & 11.1  & 10.0 \\
    MPL+utt. filter (T=0.95) & 12.8  & 11.7  & 11.2  & 9.9 \\
    APL   & \textbf{12.2} & \textbf{11.3} & \textbf{10.7} & \textbf{9.2} \\
    \bottomrule
    \end{tabular}%
  \label{tab:utt_filter}%
\end{table}%

We can observe that utterance-level filtering fails to boost the performance of MPL on the TED dataset. The reason is that although it filters out noisy pseudo-labels and improves the quality of selected pseudo-labels, it also reduces the quantity of valuable pseudo-labels. This trade-off leads to limited performance improvement if the unlabeled dataset is relatively easy \cite{zhang2020pushing} and is even harmful in some cases \cite{chen2021investigation}. In contrast, APL works like token-level filtering, where incorrect tokens are filtered out instead of entire utterances with incorrect tokens. Ideally, it can preserve all valuable tokens while excluding all harmful tokens in pseudo-labels. Therefore, it breaks the dilemma in utterance-level filtering.

In the following sections, we show detailed experiments to analyze the effectiveness of each component in APL.

\subsection{Analysis of ATC}
In this section, we compare variants and optimization schedules of ATC. Two variants (ATC-R, ATC-A) have been introduced in \autoref{sec:ATC}. We describe two optimization schedules as follows.

\begin{figure}[htbp]
    \centering
	\includegraphics[width=0.6\columnwidth]{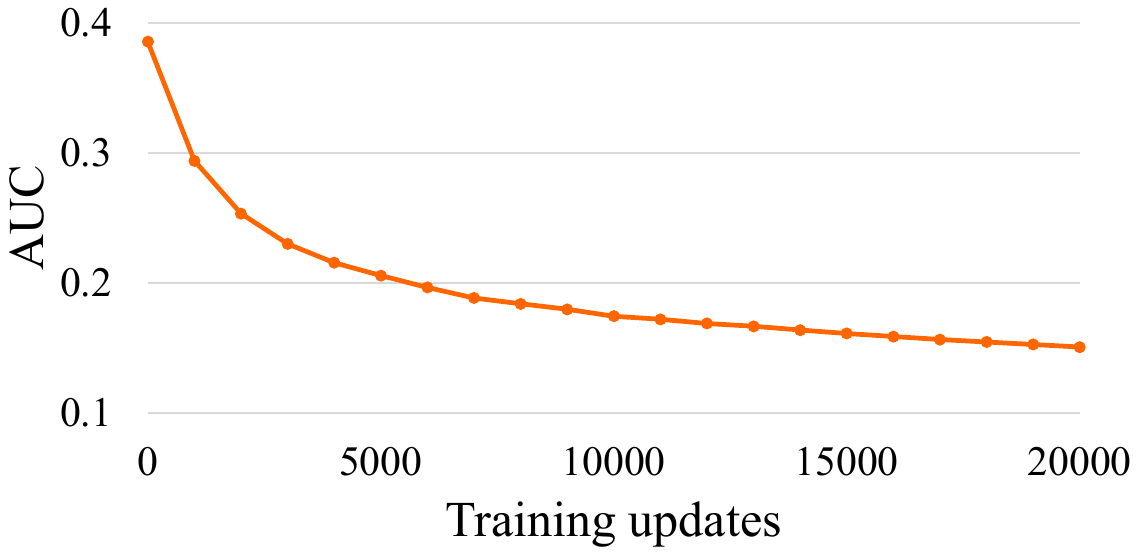}
	\caption{AUC curve on the TED training set.}
	\label{fig:AUC}
\end{figure}

The straightforward practice is to apply ATC in all training updates, which we refer to as one-step optimization. However, such practice could be suboptimal given the observation that the error detection performance is degraded during training. Specifically, we use the area under the precision-recall curve (AUC) to measure the error detection performance and treat incorrect tokens as positives when calculating the precision-recall curve. The trend of AUC on the unlabeled training set is plotted in \autoref{fig:AUC}. 
We can see that the AUC value continues to drop during training, illustrating the degradation of the error detection performance. This phenomenon is because training on erroneous pseudo-labels will mess up the confidence estimation. Consequently, in the later training updates, there will be lots of false alarm detection, i.e., correct tokens are detected as incorrect. If we create the ATC label graph accordingly, the performance will not be satisfactory. 

To tackle this issue, we can adopt a two-step optimization schedule to use ATC in the early updates and switch it to CTC in the later updates. To this end, we have introduced two optimization schedules (one- and two-step optimization).

\begin{table}[htbp]
  \centering
  \caption{Comparison of ATC variants and optimization schedules.}
    \begin{tabular}{l|cc|cc}
    \toprule
    \multicolumn{1}{l}{\multirow{3}[6]{*}{ATC varaint}} & \multicolumn{4}{c}{TED} \\
\cmidrule{2-5}    \multicolumn{1}{l}{} & \multicolumn{2}{c}{w/o LM} & \multicolumn{2}{c}{w/ LM} \\
\cmidrule(lr){2-3} \cmidrule(lr){4-5}    \multicolumn{1}{l}{} & dev   & \multicolumn{1}{c}{test} & dev   & \multicolumn{1}{c}{test} \\
    \midrule
    \multicolumn{5}{l}{\textit{One-step optimization}} \\
    ATC-A & 12.3  & 11.4  & 10.8  & 9.5 \\
    ATC-R & 13.2  & 12.2  & 10.8  & 9.6 \\
    \midrule
    \multicolumn{5}{l}{\textit{Two-step optimization}} \\
    ATC-A & 12.3  & \textbf{11.2} & 11.0  & 9.7 \\
    ATC-R & \textbf{12.2} & 11.3 & \textbf{10.7} & \textbf{9.2} \\
    \bottomrule
    \end{tabular}%
  \label{tab:compare_variant}%
\end{table}%

\begin{figure}[htbp]
    \centering
	\includegraphics[width=0.6\columnwidth]{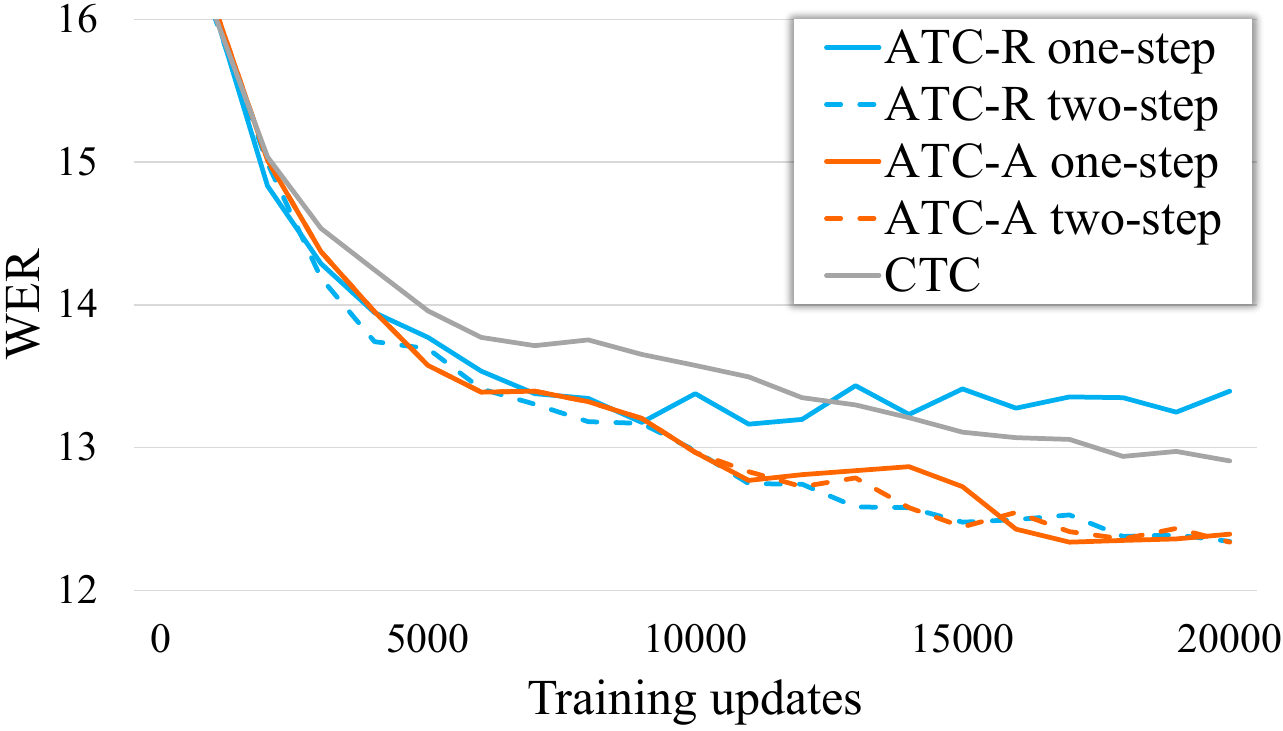}
	\caption{WER curves of different ATC variants and optimization schedules on TED development set. LM is not used in the evaluation.}
	\label{fig:analysis_variant}
\end{figure}

The performance comparison of two variants and two optimization schedules of ATC are shown in \autoref{tab:compare_variant}. ATC-R with the two-step optimization realizes the best performance. And when we switch the optimization schedule to the one-step optimization, the performance of ATC-R is clearly degraded. In contrast, two optimization schedules have similar results on the ATC-A variant. The reason is that ATC-A is designed to be more robust to the false alarm detection issue. Thus the two-step optimization designed to address the same issue has lesser benefit.
For detailed analysis, we draw the WER curves of the above combinations in \autoref{fig:analysis_variant}.

We can observe that the WER of ATC-R with one-step optimization increases in the later updates due to the degradation of confidence estimation and error detection ability. In comparison, ATC-A with one-step optimization does not show this phenomenon due to its robustness against false alarm detection.
Nonetheless, when two-step optimization is applied, both variants have reasonable WER curves and similar performance. Since ATC-R is more straightforward and has the best results when combined with two-step optimization, we use it as the default ATC method.

\begin{figure}[htbp]
    \centering
    \includegraphics[width=0.6\columnwidth]{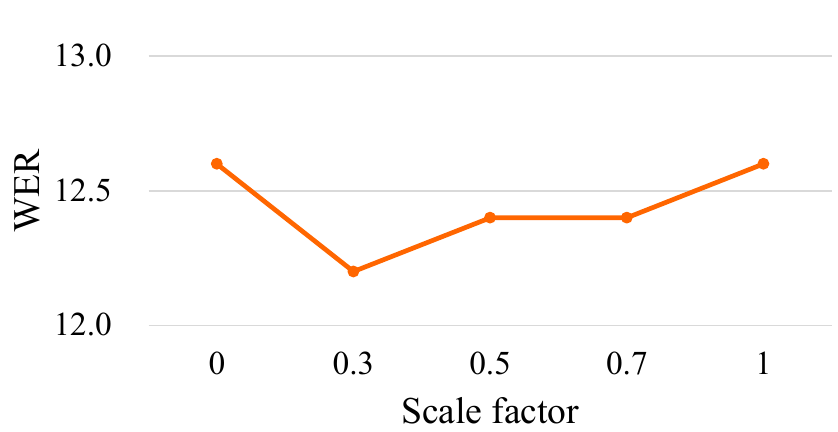}
    \caption{WER of APL (ATC-R variant) with different scale factors.}
    \label{fig:analysis_scale}

\end{figure}

Finally, we compare different scale factor $\eta$ for the probability of $*$ token. As shown in \autoref{fig:analysis_scale}, the best performance is achieved with the scale factor of 0.3. When the scale factor is 1, i.e., no scale is applied on the probability of $*$, the WER is clearly increased. In another extreme condition where the scale factor is 0, which means we delete the arcs of incorrect tokens, the performance is also not satisfactory.

\subsection{Effectiveness of Contrastive CTC}

In this section, we analyze how contrastive CTC affects the behavior of the seed model and the performance of APL. 

\begin{table}[htbp]
  \centering
  \caption{The performance of the seed model w/ and w/o contrastive learning on the TED training set.}
    \begin{tabular}{lcccc}
    \toprule
    \multirow{2}[4]{*}{Loss function} & \multirow{2}[4]{*}{WER} & \multicolumn{2}{c}{Average confidence} & \multirow{2}[4]{*}{AUC} \\
\cmidrule{3-4}          &       & Correct & Incorrect &  \\
    \midrule
    CTC   & 22.8  & 0.948 & 0.772 & 0.331 \\
    Contrastive CTC & 23.0  & 0.927 & 0.670 & 0.386 \\
    \bottomrule
    \end{tabular}%
  \label{tab:seed_contrastive}%
\end{table}%

As shown in \autoref{tab:seed_contrastive}, contrastive CTC does not decrease the WER directly. Instead, it significantly decreases the average confidence scores of incorrect tokens, thus widening the confidence distance between correct and incorrect tokens. Consequently, the AUC score of contrastive CTC is higher, illustrating the improvement in error detection.

\begin{table}[htbp]
  \centering
  \caption{WER of the final model w/ and w/o contrastive learning.}
    \begin{tabular}{l|cc|cc}
    \toprule
    \multicolumn{1}{l}{\multirow{3}[6]{*}{Loss function}} & \multicolumn{4}{c}{TED} \\
\cmidrule{2-5}    \multicolumn{1}{l}{} & \multicolumn{2}{c}{w/o LM} & \multicolumn{2}{c}{w/ LM} \\
\cmidrule(lr){2-3} \cmidrule(lr){4-5}    \multicolumn{1}{l}{} & dev   & \multicolumn{1}{c}{test} & dev   & test \\
    \midrule
    CTC   & 12.8  & 11.4  & 11.0  & 9.3 \\
    Contrastive CTC & \textbf{12.2} & \textbf{11.3} & \textbf{10.7} & \textbf{9.2} \\
    \bottomrule
    \end{tabular}%
  \label{tab:final_contrastive}%
\end{table}%

Better error detection ability can lead to a higher quality label graph of ATC, resulting in better performance. As shown in \autoref{tab:final_contrastive}, APL performs better when the seed model is trained with contrastive CTC.

\subsection{Effectiveness of Automatic Thresholding}

\begin{figure*}[htbp]
    \centering
	\subfloat[TED] 
	{ \label{fig:analysis_threshold_ted}
		\includegraphics[width=0.6\columnwidth]{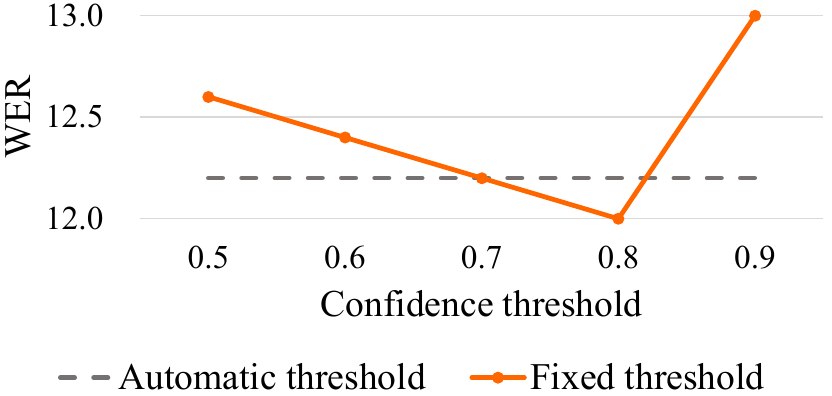}
	}\quad
	\subfloat[SWBD] 
	{ \label{fig:analysis_threshold_swbd}
		\includegraphics[width=0.6\columnwidth]{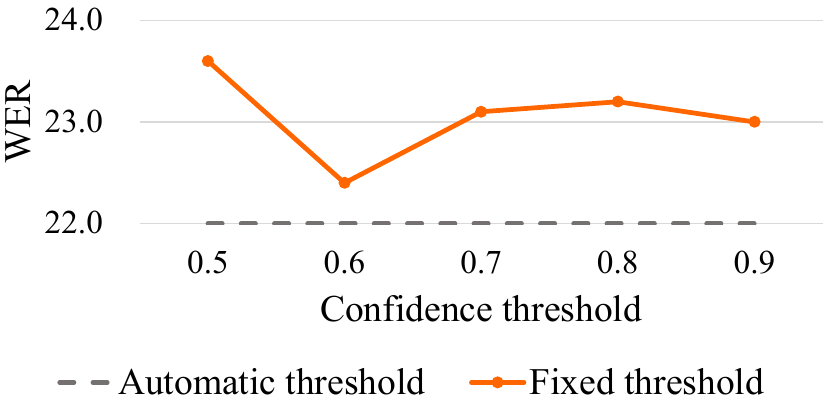}
	}\quad
	\subfloat[aidatatang] 
	{ \label{fig:analysis_threshold_aidatatang}
		\includegraphics[width=0.6\columnwidth]{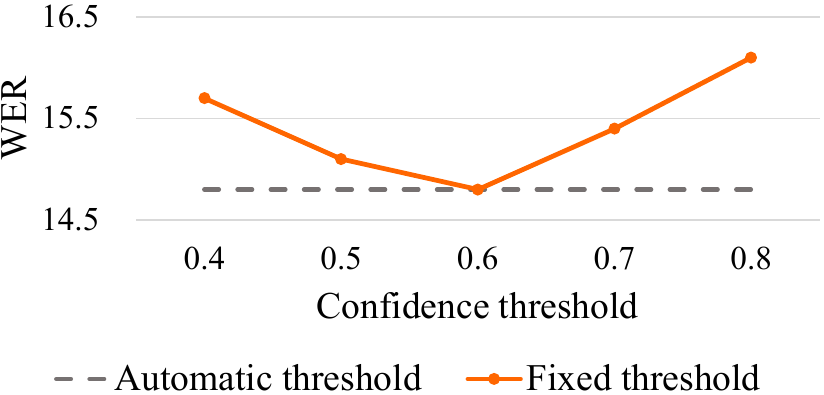}
	}
	\caption{Comparison of the proposed automatic threshold and various fixed thresholds.} 
	\label{fig:confidence}
\end{figure*}

In this section, we examine the effectiveness of automatic thresholding. Automatic thresholding is proposed to avoid the manual tuning of thresholds. Therefore, we compare the automatic threshold with various fixed thresholds. As shown in \autoref{fig:confidence}, the optimal fixed threshold varies among datasets. Hence, using a fixed threshold for all datasets is not optimal, and manually tuning thresholds for each dataset is required.
In contrast, automatic thresholding eliminates the requirement of tuning thresholds while performing similarly with or even better than the best manually tuned fixed threshold on various datasets.

\begin{table*}[htbp]
  \centering
  \caption{WER of APL w/ and w/o relative correction.}
    \begin{tabular}{l|cc|cc|ccc|ccc|cc|cc}
    \toprule
    \multicolumn{1}{l}{\multirow{3}[6]{*}{Method}} & \multicolumn{4}{c}{TED}       & \multicolumn{6}{c}{SWBD}                      & \multicolumn{4}{c}{aidatatang} \\
\cmidrule(lr){2-5}  \cmidrule(lr){6-11}  \cmidrule(lr){12-15}     \multicolumn{1}{l}{} & \multicolumn{2}{c}{w/o LM} & \multicolumn{2}{c}{w/ LM} & \multicolumn{3}{c}{w/o LM} & \multicolumn{3}{c}{w/ LM} & \multicolumn{2}{c}{w/o LM} & \multicolumn{2}{c}{w/ LM} \\
\cmidrule(lr){2-3} \cmidrule(lr){4-5}  \cmidrule(lr){6-8} \cmidrule(lr){9-11} \cmidrule(lr){12-13} \cmidrule(lr){14-15}     \multicolumn{1}{l}{} & dev   & \multicolumn{1}{c}{test} & dev   & \multicolumn{1}{c}{test} & RT03  & H-SB  & \multicolumn{1}{c}{H-CH} & RT03  & H-SB  & \multicolumn{1}{c}{H-CH} & dev   & \multicolumn{1}{c}{test} & dev   & test \\
    \midrule
    w/o relative & \textbf{12.1}  & \textbf{11.2}  & \textbf{10.5}  & \textbf{9.2}   & 22.8  & \textbf{16.0}  & 22.7  & 19.3  & 13.3  & 19.3  & 15.1  & 15.7  & 13.7  & 14.5 \\
    w/ relative & 12.2 & 11.3 & 10.7 & \textbf{9.2} & \textbf{22.0} & \textbf{16.0} & \textbf{22.2} & \textbf{18.5} & \textbf{13.3} & \textbf{19.1} & \textbf{14.8} & \textbf{15.4} & \textbf{13.5} & \textbf{14.3} \\
    \bottomrule
    \end{tabular}%
  \label{tab:ablation_relative}%
\end{table*}%

Moreover, a unique design in automatic thresholding is the relative correction. We conduct the corresponding ablation study and show the results in \autoref{tab:ablation_relative}. We can observe that the relative correction keeps a similar performance on the slightly mismatched condition (TED) and clearly improves the performance on the severely mismatched condition (SWBD), proving its effectiveness in remedying the domain gap of unlabeled and labeled data.

\section{Conclusion}
\label{sec:conclusion}
In this work, we propose to address the issue of erroneous pseudo-labels in semi-supervised learning from the perspective of training objective. The core method is a generalized CTC loss function named ATC that can deal with noisy pseudo-labels. Furthermore, since ATC's performance relies on the quality of error detection with confidence scores, we introduce contrastive CTC to widen the confidence gap between the correctly and incorrectly predicted tokens, thus improving error detection. Additionally, we design an automatic thresholding method to avoid the expensive manual tuning of the confidence threshold. The automatic thresholding method achieves comparable or even better than a carefully tuned confidence threshold.
Finally, we formulate the APL approach for semi-supervised ASR with all the above innovations. Experiments on various datasets demonstrate that APL outperforms existing pseudo-labeling approaches, realizing 23.4\% and 7.2\% average relative WER reduction compared with IPL and MPL-based semi-supervised ASR, respectively. 

We have demonstrated the effectiveness of the proposed approach on CTC models. Since the ideas in the proposed approach are not model-specific, a future direction is to apply the ideas of tackling noisy pseudo-labels in the loss function, contrastive learning for better confidence score, and automatic thresholding on AED~\cite{chan2016listen} and RNN-T~\cite{graves2012sequence} based E2E-ASR models.

\section*{Acknowledgment}

This work is partially supported by the National Key R\&D Program of China (2022ZD0116103), the Youth Innovation Promotion Association, Chinese Academy of Sciences, and the Frontier Exploration Project Independently Deployed by Institute of Acoustics, Chinese Academy of Sciences under Grant QYTS202011. 

\ifCLASSOPTIONcaptionsoff
  \newpage
\fi

\bibliographystyle{IEEEtran}
\bibliography{./main}

\begin{thebibliography}{10}
\providecommand{\url}[1]{#1}
\csname url@samestyle\endcsname
\providecommand{\newblock}{\relax}
\providecommand{\bibinfo}[2]{#2}
\providecommand{\BIBentrySTDinterwordspacing}{\spaceskip=0pt\relax}
\providecommand{\BIBentryALTinterwordstretchfactor}{4}
\providecommand{\BIBentryALTinterwordspacing}{\spaceskip=\fontdimen2\font plus
\BIBentryALTinterwordstretchfactor\fontdimen3\font minus
  \fontdimen4\font\relax}
\providecommand{\BIBforeignlanguage}[2]{{%
\expandafter\ifx\csname l@#1\endcsname\relax
\typeout{** WARNING: IEEEtran.bst: No hyphenation pattern has been}%
\typeout{** loaded for the language `#1'. Using the pattern for}%
\typeout{** the default language instead.}%
\else
\language=\csname l@#1\endcsname
\fi
#2}}
\providecommand{\BIBdecl}{\relax}
\BIBdecl

\bibitem{li2021recent}
J.~Li, ``Recent advances in end-to-end automatic speech recognition,''
  \emph{APSIPA Transactions on Signal and Information Processing}, 2021.

\bibitem{cheng2022eteh}
G.~Cheng, H.~Miao, R.~Yang, K.~Deng, and Y.~Yan, ``Eteh: Unified
  attention-based end-to-end asr and kws architecture,'' \emph{IEEE/ACM
  Transactions on Audio, Speech, and Language Processing}, vol.~30, pp.
  1360--1373, 2022.

\bibitem{peddinti2015time}
V.~Peddinti, D.~Povey, and S.~Khudanpur, ``A time delay neural network
  architecture for efficient modeling of long temporal contexts,'' in
  \emph{Sixteenth annual conference of the international speech communication
  association}, 2015.

\bibitem{povey2018semi}
D.~Povey, G.~Cheng, Y.~Wang, K.~Li, H.~Xu, M.~Yarmohammadi, and S.~Khudanpur,
  ``Semi-orthogonal low-rank matrix factorization for deep neural networks.''
  in \emph{Interspeech}, 2018, pp. 3743--3747.

\bibitem{gulati2020conformer}
A.~Gulati, J.~Qin, C.-C. Chiu, N.~Parmar, Y.~Zhang, J.~Yu, W.~Han, S.~Wang,
  Z.~Zhang, Y.~Wu, and R.~Pang, ``{Conformer: Convolution-augmented Transformer
  for Speech Recognition},'' in \emph{Proc. Interspeech 2020}, 2020, pp.
  5036--5040.

\bibitem{baevski2020wav2vec}
A.~Baevski, Y.~Zhou, A.~Mohamed, and M.~Auli, ``Wav2vec 2.0: A framework for
  self-supervised learning of speech representations,'' \emph{Advances in
  Neural Information Processing Systems}, vol.~33, 2020.

\bibitem{hsu2021hubert}
W.-N. Hsu, Y.-H.~H. Tsai, B.~Bolte, R.~Salakhutdinov, and A.~Mohamed, ``Hubert:
  How much can a bad teacher benefit asr pre-training?'' in \emph{ICASSP
  2021-2021 IEEE International Conference on Acoustics, Speech and Signal
  Processing (ICASSP)}.\hskip 1em plus 0.5em minus 0.4em\relax IEEE, 2021, pp.
  6533--6537.

\bibitem{higuchi2021momentum}
Y.~Higuchi, N.~Moritz, J.~L. Roux, and T.~Hori, ``{Momentum Pseudo-Labeling for
  Semi-Supervised Speech Recognition},'' in \emph{Proc. Interspeech 2021},
  2021, pp. 726--730.

\bibitem{park2020improved}
D.~S. Park, Y.~Zhang, Y.~Jia, W.~Han, C.-C. Chiu, B.~Li, Y.~Wu, and Q.~V. Le,
  ``Improved noisy student training for automatic speech recognition,''
  \emph{Proc. Interspeech 2020}, pp. 2817--2821, 2020.

\bibitem{fu2021incremental}
L.~Fu, X.~Li, L.~Zi, Z.~Zhang, Y.~Wu, X.~He, and B.~Zhou, ``Incremental
  learning for end-to-end automatic speech recognition,'' in \emph{2021 IEEE
  Automatic Speech Recognition and Understanding Workshop (ASRU)}.\hskip 1em
  plus 0.5em minus 0.4em\relax IEEE, 2021, pp. 320--327.

\bibitem{likhomanenko2020slimipl}
T.~Likhomanenko, Q.~Xu, J.~Kahn, G.~Synnaeve, and R.~Collobert, ``{slimIPL:
  Language-Model-Free Iterative Pseudo-Labeling},'' in \emph{Proc. Interspeech
  2021}, 2021, pp. 741--745.

\bibitem{kahn2020self}
J.~Kahn, A.~Lee, and A.~Hannun, ``Self-training for end-to-end speech
  recognition,'' in \emph{ICASSP 2020-2020 IEEE International Conference on
  Acoustics, Speech and Signal Processing (ICASSP)}.\hskip 1em plus 0.5em minus
  0.4em\relax IEEE, 2020, pp. 7084--7088.

\bibitem{khurana2021unsupervised}
S.~Khurana, N.~Moritz, T.~Hori, and J.~Le~Roux, ``Unsupervised domain
  adaptation for speech recognition via uncertainty driven self-training,'' in
  \emph{ICASSP 2021-2021 IEEE International Conference on Acoustics, Speech and
  Signal Processing (ICASSP)}.\hskip 1em plus 0.5em minus 0.4em\relax IEEE,
  2021, pp. 6553--6557.

\bibitem{xu2020iterative}
Q.~Xu, T.~Likhomanenko, J.~Kahn, A.~Hannun, G.~Synnaeve, and R.~Collobert,
  ``Iterative pseudo-labeling for speech recognition,'' \emph{Proc. Interspeech
  2020}, pp. 1006--1010, 2020.

\bibitem{manohar2021kaizen}
V.~Manohar, T.~Likhomanenko, Q.~Xu, W.-N. Hsu, R.~Collobert, Y.~Saraf,
  G.~Zweig, and A.~Mohamed, ``Kaizen: Continuously improving teacher using
  exponential moving average for semi-supervised speech recognition,'' in
  \emph{2021 IEEE Automatic Speech Recognition and Understanding Workshop
  (ASRU)}, 2021, pp. 518--525.

\bibitem{chen2020semi}
Y.~Chen, W.~Wang, and C.~Wang, ``Semi-supervised asr by end-to-end
  self-training,'' \emph{Proc. Interspeech 2020}, pp. 2787--2791, 2020.

\bibitem{zhang2020pushing}
Y.~Zhang, J.~Qin, D.~S. Park, W.~Han, C.-C. Chiu, R.~Pang, Q.~V. Le, and Y.~Wu,
  ``Pushing the limits of semi-supervised learning for automatic speech
  recognition,'' \emph{NeurIPS SAS 2020 Workshop}, 2020.

\bibitem{zhu2022boosting}
H.~Zhu, G.~Cheng, J.~Wang, W.~Hou, P.~Zhang, and Y.~Yan, ``Boosting
  cross-domain speech recognition with self-supervision,'' \emph{IEEE/ACM
  Transactions on Audio, Speech, and Language Processing}, pp. 1--15, 2023.

\bibitem{berthelot2019mixmatch}
D.~Berthelot, N.~Carlini, I.~Goodfellow, N.~Papernot, A.~Oliver, and C.~A.
  Raffel, ``Mixmatch: A holistic approach to semi-supervised learning,''
  \emph{Advances in neural information processing systems}, vol.~32, 2019.

\bibitem{chen21c_interspeech}
Z.~Chen, A.~Rosenberg, Y.~Zhang, H.~Zen, M.~Ghodsi, Y.~Huang, J.~Emond,
  G.~Wang, B.~Ramabhadran, and P.~J. Moreno, ``{Semi-Supervision in ASR:
  Sequential MixMatch and Factorized TTS-Based Augmentation},'' in \emph{Proc.
  Interspeech 2021}, 2021, pp. 736--740.

\bibitem{graves2006connectionist}
A.~Graves, S.~Fern{\'a}ndez, F.~Gomez, and J.~Schmidhuber, ``Connectionist
  temporal classification: labelling unsegmented sequence data with recurrent
  neural networks,'' in \emph{Proceedings of the 23rd international conference
  on Machine learning}, 2006, pp. 369--376.

\bibitem{zhang2014semi}
P.~Zhang, Y.~Liu, and T.~Hain, ``Semi-supervised dnn training in meeting
  recognition,'' in \emph{2014 IEEE Spoken Language Technology Workshop
  (SLT)}.\hskip 1em plus 0.5em minus 0.4em\relax IEEE, 2014, pp. 141--146.

\bibitem{li2021confidence}
Q.~Li, D.~Qiu, Y.~Zhang, B.~Li, Y.~He, P.~C. Woodland, L.~Cao, and T.~Strohman,
  ``Confidence estimation for attention-based sequence-to-sequence models for
  speech recognition,'' in \emph{ICASSP 2021-2021 IEEE International Conference
  on Acoustics, Speech and Signal Processing (ICASSP)}.\hskip 1em plus 0.5em
  minus 0.4em\relax IEEE, 2021, pp. 6388--6392.

\bibitem{zheng2021rectifying}
Z.~Zheng and Y.~Yang, ``Rectifying pseudo label learning via uncertainty
  estimation for domain adaptive semantic segmentation,'' \emph{International
  Journal of Computer Vision}, vol. 129, no.~4, pp. 1106--1120, 2021.

\bibitem{gal2016dropout}
Y.~Gal and Z.~Ghahramani, ``Dropout as a bayesian approximation: Representing
  model uncertainty in deep learning,'' in \emph{international conference on
  machine learning}.\hskip 1em plus 0.5em minus 0.4em\relax PMLR, 2016, pp.
  1050--1059.

\bibitem{cai2021w}
X.~Cai, J.~Yuan, Y.~Bian, G.~Xun, J.~Huang, and K.~Church, ``W-ctc: a
  connectionist temporal classification loss with wild cards,'' in
  \emph{International Conference on Learning Representations}, 2021.

\bibitem{pratapstar}
V.~Pratap, A.~Hannun, G.~Synnaeve, and R.~Collobert, ``Star temporal
  classification: Sequence modeling with partially labeled data,''
  \emph{Advances in Neural Information Processing Systems}, vol.~35, pp.
  13\,392--13\,403, 2022.

\bibitem{dufraux2019lead2gold}
A.~Dufraux, E.~Vincent, A.~Hannun, A.~Brun, and M.~Douze, ``Lead2gold: Towards
  exploiting the full potential of noisy transcriptions for speech
  recognition,'' in \emph{2019 IEEE Automatic Speech Recognition and
  Understanding Workshop (ASRU)}.\hskip 1em plus 0.5em minus 0.4em\relax IEEE,
  2019, pp. 78--85.

\bibitem{do2021multiple}
C.-T. Do, R.~Doddipatla, and T.~Hain, ``Multiple-hypothesis ctc-based
  semi-supervised adaptation of end-to-end speech recognition,'' in
  \emph{ICASSP 2021-2021 IEEE International Conference on Acoustics, Speech and
  Signal Processing (ICASSP)}.\hskip 1em plus 0.5em minus 0.4em\relax IEEE,
  2021, pp. 6978--6982.

\bibitem{moritz2021semi}
N.~Moritz, T.~Hori, and J.~Le~Roux, ``Semi-supervised speech recognition via
  graph-based temporal classification,'' in \emph{ICASSP 2021-2021 IEEE
  International Conference on Acoustics, Speech and Signal Processing
  (ICASSP)}.\hskip 1em plus 0.5em minus 0.4em\relax IEEE, 2021, pp. 6548--6552.

\bibitem{manohar2018semi}
V.~Manohar, H.~Hadian, D.~Povey, and S.~Khudanpur, ``Semi-supervised training
  of acoustic models using lattice-free mmi,'' in \emph{2018 IEEE international
  conference on acoustics, speech and signal processing (ICASSP)}.\hskip 1em
  plus 0.5em minus 0.4em\relax IEEE, 2018, pp. 4844--4848.

\bibitem{mohri2002weighted}
M.~Mohri, F.~Pereira, and M.~Riley, ``Weighted finite-state transducers in
  speech recognition,'' \emph{Computer Speech \& Language}, vol.~16, no.~1, pp.
  69--88, 2002.

\bibitem{hannun2020differentiable}
A.~Hannun, V.~Pratap, J.~Kahn, and W.-N. Hsu, ``Differentiable weighted
  finite-state transducers,'' \emph{arXiv preprint arXiv:2010.01003}, 2020.

\bibitem{miao2015eesen}
Y.~Miao, M.~Gowayyed, and F.~Metze, ``Eesen: End-to-end speech recognition
  using deep rnn models and wfst-based decoding,'' in \emph{2015 IEEE Workshop
  on Automatic Speech Recognition and Understanding (ASRU)}.\hskip 1em plus
  0.5em minus 0.4em\relax IEEE, 2015, pp. 167--174.

\bibitem{laptev22_interspeech}
A.~Laptev, S.~Majumdar, and B.~Ginsburg, ``{CTC Variations Through New WFST
  Topologies},'' in \emph{Proc. Interspeech 2022}, 2022, pp. 1041--1045.

\bibitem{wang2022freematch}
Y.~Wang, H.~Chen, Q.~Heng, W.~Hou, Y.~Fan, Z.~Wu, J.~Wang, M.~Savvides,
  T.~Shinozaki, B.~Raj \emph{et~al.}, ``Freematch: Self-adaptive thresholding
  for semi-supervised learning,'' in \emph{The Eleventh International
  Conference on Learning Representations}, 2022.

\bibitem{panayotov2015librispeech}
V.~Panayotov, G.~Chen, D.~Povey, and S.~Khudanpur, ``Librispeech: an asr corpus
  based on public domain audio books,'' in \emph{2015 IEEE international
  conference on acoustics, speech and signal processing (ICASSP)}.\hskip 1em
  plus 0.5em minus 0.4em\relax IEEE, 2015, pp. 5206--5210.

\bibitem{hernandez2018ted}
F.~Hernandez, V.~Nguyen, S.~Ghannay, N.~Tomashenko, and Y.~Esteve, ``Ted-lium
  3: twice as much data and corpus repartition for experiments on speaker
  adaptation,'' in \emph{International conference on speech and
  computer}.\hskip 1em plus 0.5em minus 0.4em\relax Springer, 2018, pp.
  198--208.

\bibitem{godfrey1993switchboard}
J.~Godfrey and E.~Holliman, ``Switchboard-1 release 2 ldc97s62,'' \emph{Web
  Download. Philadelphia: Linguistic Data Consortium}, 1993.

\bibitem{aishell_2017}
H.~Bu, J.~Du, X.~Na, B.~Wu, and H.~Zheng, ``Aishell-1: An open-source mandarin
  speech corpus and a speech recognition baseline,'' in \emph{2017 20th
  Conference of the Oriental Chapter of the International Coordinating
  Committee on Speech Databases and Speech I/O Systems and Assessment
  (O-COCOSDA)}, 2017, pp. 1--5.

\bibitem{aidatatang}
{Beijing DataTang Technology Co., Ltd}, ``aidatatang\_200zh, a free chinese
  mandarin speech corpus.''

\bibitem{rt03}
e.~a. Fiscus, Jonathan~G., ``2003 nist rich transcription evaluation data
  ldc2007s10,'' \emph{Web Download. Philadelphia: Linguistic Data Consortium},
  2007.

\bibitem{eval2000}
{Linguistic Data Consortium}, ``2000 hub5 english evaluation speech
  ldc2002s09,'' \emph{Web Download. Philadelphia: Linguistic Data Consortium},
  2002.

\bibitem{ott2019fairseq}
M.~Ott, S.~Edunov, A.~Baevski, A.~Fan, S.~Gross, N.~Ng, D.~Grangier, and
  M.~Auli, ``fairseq: A fast, extensible toolkit for sequence modeling,'' in
  \emph{Proceedings of NAACL-HLT 2019: Demonstrations}, 2019.

\bibitem{park2019specaugment}
D.~S. Park, W.~Chan, Y.~Zhang, C.-C. Chiu, B.~Zoph, E.~D. Cubuk, and Q.~V. Le,
  ``Specaugment: A simple data augmentation method for automatic speech
  recognition,'' \emph{Proc. Interspeech 2019}, pp. 2613--2617, 2019.

\bibitem{chen2021investigation}
I.-F. Chen, B.~King, and J.~Droppo, ``Investigation of training label error
  impact on rnn-t,'' \emph{arXiv preprint arXiv:2112.00350}, 2021.

\bibitem{chan2016listen}
W.~Chan, N.~Jaitly, Q.~Le, and O.~Vinyals, ``Listen, attend and spell: A neural
  network for large vocabulary conversational speech recognition,'' in
  \emph{2016 IEEE international conference on acoustics, speech and signal
  processing (ICASSP)}.\hskip 1em plus 0.5em minus 0.4em\relax IEEE, 2016, pp.
  4960--4964.

\bibitem{graves2012sequence}
A.~Graves, ``Sequence transduction with recurrent neural networks,'' \emph{ICML
  Workshop on Representation Learning}, 2012.

\end{thebibliography}

\end{document}